\documentclass[%
reprint,
 amsmath,amssymb,
 aps,
 prl,
floatfix,
]{revtex4-1}
\usepackage{graphicx,xcolor}
\definecolor{darkblue}{RGB}{0,0,150}
\definecolor{nightblue}{RGB}{0,0,100}

\usepackage{mathrsfs,dsfont,mathtools}

\usepackage{dcolumn}
\usepackage{bm}
\usepackage[
colorlinks,
citecolor=darkblue,
linkcolor=darkblue,
urlcolor=nightblue]{hyperref}

\usepackage[english]{babel}
\usepackage[babel,kerning=true,spacing=true]{microtype}

\usepackage{feynmp-auto}

\renewcommand{\Re}[1]{\textrm{Re}\left( #1 \right)}

\newcommand{\lra}{\leftrightarrow}
\newcommand{\nd}{2\textsuperscript{nd}}

\newcommand{\ket}[1]{\left| #1 \right\rangle}
\newcommand{\bra}[1]{\left\langle #1 \right|}
\newcommand{\braket}[3]{\langle #1 \left| #2 \right| #3  \rangle}
\newcommand{\mA}{\mathcal{A}}

\definecolor{DarkRed}{RGB}{100,0,0}
\definecolor{DarkBlue}{RGB}{000,0,100}

\AtBeginDocument{\renewcommand{\natexlab}[1]{#1}}
\begin{document}

\title{
General nonlinear Hall current in magnetic insulators beyond the quantum anomalous Hall effect
}

\author{Daniel Kaplan}
\author{Tobias Holder}
\author{Binghai Yan}
\email{binghai.yan@weizmann.ac.il}
\affiliation{Department of Condensed Matter Physics,
Weizmann Institute of Science,
Rehovot 7610001, Israel}

\date{\today}
\begin{abstract}
Can a generic magnetic insulator exhibit a Hall current? The quantum anomalous Hall effect (QAHE) is one example of an insulating bulk carrying a quantized Hall conductivity and other insulators (with zero Chern number) present zero Hall conductance in the linear response regime. Here, we find that a  general magnetic insulator possesses a nonlinear Hall conductivity quadratic to the electric field if the system breaks inversion symmetry. This conductivity originates from an induced orbital magnetization due to virtual interband transitions. We identify three contributions to the wavepacket motion,  a velocity shift, a positional shift, and a Berry curvature renormalization. In contrast to the crystalline solid, we find that this nonlinear Hall conductivity vanishes for Landau levels of a 2D electron gas, indicating a fundamental difference between the QAHE and the Integer quantum Hall effect.

\end{abstract}

\maketitle

\emph{Introduction.---}
Understanding electric conduction of insulators is fundamental to condensed matter physics. For example, 
the quantum Hall effect is a unique realization of a 2D topological insulating phase of matter, with distinct experimental signatures \cite{vonKlitzing1980,vonKlitzing1986,Senthil2015,Hansson2017}, notably a quantized Hall conductance $\sigma^{xy}$ which adheres to the quantized value $\frac{e^2}{h}$ with astonishing precision, up to at least $10^{-10}$ \cite{Schopfer2007,Poirier2009}. 
It has been known since the early days of the quantum Hall effect~\cite{Thouless1982} that the quantization of $\sigma_{xy}$ is related to the Berry curvature in a periodic system, with the robustness of the quantization discussed in several works \cite{Altshuler1980,Avron1983,Zala2001}. 
As a close cousin, the quantum anomalous Hall effect (QAHE) refers to the appearance of a quantized Hall conductivity in 2D systems even in the absence of a magnetic field \cite{Liu2016b,He2018,Chang2022}. 
First proposed by Haldane \cite{Haldane1988}, the QAHE requires the breaking of time-reversal symmetry (TRS) in the crystal system characterized by a Chern number $C_N$ for the occupied bands. 
Consequently, a calculation at linear order yields for the Hall conductivity $\sigma^{xy} = C_N\frac{e^2}{h}$~\cite{Xiao2010,Nagaosa2010}. 
The QAHE has been experimentally realized in several systems, notably magnetically doped thin-films of topological insulators \cite{Chang2013,Chang2015high}, stochiometric magnetic topological insulators~\cite{Deng2020}, and recently in Moir\'e superlattices~\cite{Serlin2020, Li2021}.
However, in contrast to the quantum Hall effect, careful experiments on the QAHE find a less precisely quantized Hall conductivity,
with precision of $0.01\%$~\cite{Chang2015high} and $0.1\%$~\cite{Serlin2020} respectively.

\begin{figure*}
    \centering
    \includegraphics[width=0.7\textwidth]{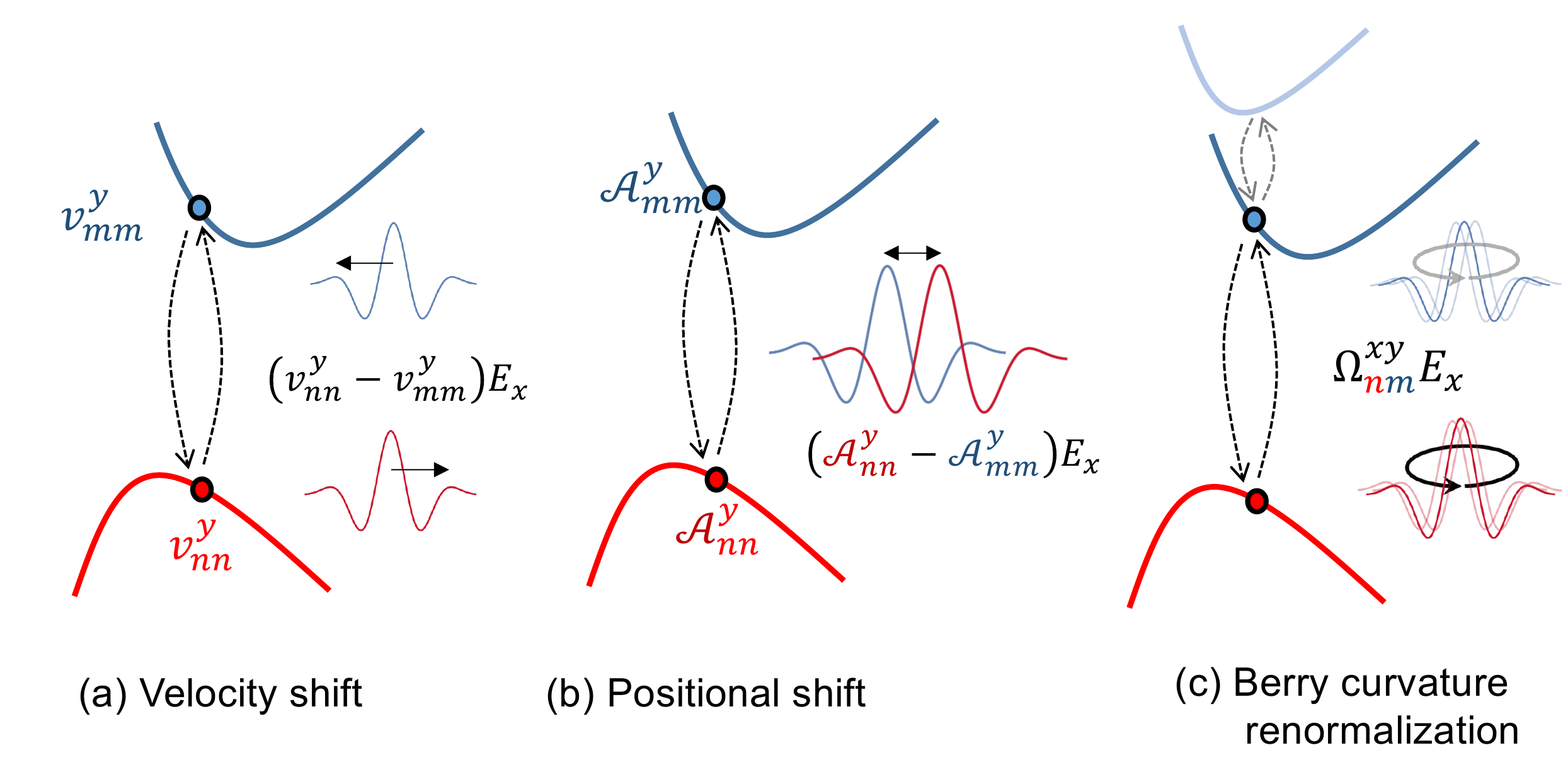}
    \caption{\textbf{Illustration of corrections to the AHE conductivity in the presence of an electric field ($E_x$)}. (a) Velocity shift of the wave packet due to transition  between the occupied ($n$) and unoccupied ($m$) bands. (b) Positional shift of a wavepacket due to interband transitions. (c) Berry curvature renormalization by the third bands. All three contributions (see Eqs.~\ref{eq:i1},~\ref{eq:i2} and \ref{eq:fconductivity}), which are linear in $E_x$, are non-zero for a generic multiband dispersion which breaks both inversion and time-reversal symmetries.}
    \label{fig:overview}
\end{figure*}

While it is well established that the Hall conductivity is exactly quantized at linear order~\cite{Thouless1982}, in this letter we find that an intrinsic nonlinear conductivity can appear at finite bias for generic magnetic insulators.
These nonlinear effects are due to virtual interband transitions, which have been shown to appear beyond linear response even in insulators~\cite{Michishita2021,Kaplan2020}.
This is a testament to the fact that the Hall conductivity is a response function which can depend on the applied bias and which is not identical in physical content to the Chern number of the ground state~\cite{Holder2021a}. 
An intuitive picture of these interband transitions is shown in Fig.~\ref{fig:overview}: 
The quasiparticle response can be modified by shifts in velocity, shifts in position, and by a renormalization of the Berry curvature. In the following, we  derive the nonlinear conductivity in quantum perturbation theory, and formulate our main result as a correction to the semiclassical equations of motion.
An immediate question is why such effects are absent in the integer quantum Hall effect. The reason is that the Berry phase is entirely created by the applied magnetic field and thus extrinsic and independent of the underlying band structure, which is fully renormalized and transformed into Landau levels. Thus, no correction to the Berry phase is expected from band structure terms, and indeed as we will show, the higher-order terms considered here are rendered zero for a Landau level dispersion.


\paragraph{Theory.---}
As is well known, an applied perturbation does not usually commute with the band Hamiltonian $H_0$ and will induce interband transitions in terms of the unperturbed eigenvalues of $H_0$. Thus, a wavepacket initially centered on a single Bloch periodic state $\ket{W} = \int \mathrm{d}\mathbf{k} a_n(\mathbf{k})\ket{n\mathbf{k}}$ at $t = 0$, will evolve to be a linear combination containing contributions of many bands~\cite{Culcer2005,Chang2008}. In the Kubo formalism this is reflected in the appearance of resonant contributions which are broadened by the finite quasiparticle lifetime $\tau$.
We consider an insulator with broken inversion and time-reversal symmetries, i.~e. it holds for the dispersion $\varepsilon_n (\mathbf{k}) \neq \varepsilon_n (-\mathbf{k})$.
The uniform ($\mathbf{q} \to 0$) electric field $\mathbf{E} = \mathbf{E_0} e^{i\omega t}$ is introduced via its vector potential $\mathbf{A}(t) = \frac{\mathbf{E_0} e^{i\omega t}}{i\omega}$. 
By minimal coupling the Bloch-periodic Hamiltonian transforms as $H_0(\mathbf{k}) \to H_0(\mathbf{k}-e\mathbf{A})$. The current operator is given by $J^c = \frac{\delta H}{\delta A}$. Up to $\mathbf{A}^2$ this yields
\begin{align}
    J^c = -e v^c + e^2 \sum_{a} \mathbf{A}^a w^{ac}-\frac{e^3}{2} \sum_{a,b} \mathbf{A}^a \mathbf{A}^b u^{abc},
    \label{eq:single_particle_curr}
\end{align}
where $v^a = \partial_a H_0, w^{ab} = \partial_a \partial_b H_0, u^{abc} = \partial_a \partial_b \partial_c H_0$ and $\partial_a = \frac{\partial}{\partial_{k_a}}$.
The evaluation of the total current is then carried out using a Green's function approach \cite{Mahan1990,Jishi2013},
\begin{align}
    \langle j^c \rangle = -i \int \frac{\textrm{d}^d k}{(2\pi)^{d}} \int \frac{d\Omega}{2\pi} \mathrm{Tr}\left(J^c G^{<}(k,\Omega)\right). \label{eq:total_curr}
\end{align}
Here $G^{<}$ is the lesser Green's function \cite{Jishi2013}. The lesser Green's function is found through a solution to the Dyson equation, giving $G^{<} = (1 + G^r \Sigma^r)G^{<}_0(1 + G^a \Sigma^a)$, and $G^{r} = G_0^r (1 + \Sigma^r G^{r}),G^{a} = G_0^a (1 + \Sigma^a G^{a})$. In the usual manner \cite{Mahan1990}, Dyson's equations are solved perturbatively. Since the electromagnetic coupling is Hermitian,  $\Sigma^r = \Sigma^a = -\sum J^c \mathbf{A}^c(\omega)$, and $G^{r/a} = G_0^{r/a} \sum_{n = 0}^\infty (\Sigma^{r/a} G_0^{r/a})^n$, and correspondingly $G^{<} = \sum_{n = 0}^\infty (\Sigma^{r} G_0^{r})^n G^{<}_0 \sum_{m = 0}^\infty (\Sigma^{a} G_0^{a})^m$. The diagrammatic expansion of $G^{<}$ has recently been developed yielding the complete response at \nd~order in $\mathbf{A}$ \cite{Parker2019, Holder2020}. In the Bloch basis $\ket{n\mathbf{k}}$ the unperturbed Green's functions are $G^{r}_{0,nm}(\Omega) = \tfrac{\delta_{nm}}{\Omega - \varepsilon_n + i\tau^{-1}}$, $G^{a}_{0,nm}(\Omega) = \tfrac{\delta_{nm}}{\Omega - \varepsilon_n - i\tau^{-1}}$, $G^{<}_{0,nm}(\Omega) = 2\pi i \delta_{nm}f_{n}\delta(\Omega - \varepsilon_n)$. Here $f_n$ is the Fermi occupation factor. We begin by considering $\mathbf{A}(\omega)$ at finite frequency, and then taking the limit $\omega \to 0$. Crucially, the $\omega \to 0$ pole is avoided by retardation in the form of $\omega \to \omega + \frac{i}{\tau}$. The expansion of $G^{<}$ will contain a pole in the sum of frequencies, which is shifted by $\omega +(- \omega) \to \omega +(-\omega) + \frac{2i}{\tau}$. The result is then evaluated in the $\tau \to \infty$ limit. The expansion for the lifetime-free ($\tau^0$) contribution is detailed in the Supplementary Information~\footnote{See Supplementary Information, where the derivation of the Hall conductivity is documented and general gauge invariance is discussed.}. The full expansion for all orders of $\tau$ and the general expressions are presented in Ref.~\cite{Kaplan2022a}.
For concreteness, we present the case $\mathbf{E_0} = (E_x,0,0)$ and focus on two-dimensional systems. At order $\tau^{0}$, and up to order $\mathcal{O}(\mathbf{A}^2)$ the transverse conductivity $\sigma^{xy}$ reads,
\begin{align}
    \sigma^{xy} &=\frac{e^2}{\hbar}\!\!\!\sum_{n \in \textrm{occ.}}\! \int\!\! \frac{\mathrm{d}^2 k}{(2\pi)^2}\!
    \biggl[\Omega^{xy}_{nn} 
    +e E_x(I_1+I_2+I_3)^{xy}_{nn}\biggr]
    \label{eq:mainresult}
\end{align}
\begin{align}    
    (I_1)^{xy}_{nn}&=
    \left[\varepsilon^{-2} \mA^x, \Delta^y \mA^x\right]_{nn}
    -\left[\varepsilon^{-2} \mA^y, \Delta^x \mA^x\right]_{nn}
    \label{eq:i1}
    \\
    (I_2)^{xy}_{nn}&=
    2\left[\varepsilon^{-1} \mA^x, S^{xy}\right] - 2\left[\varepsilon^{-1} \mA^y, S^{xx}\right]
    \label{eq:i2}
    \\
    (I_3)^{xy}_{nn}&=
    i\left[\varepsilon^{-1} \mA^x, [\mA^x, \mA^y]\right]_{nn}.
     \label{eq:fconductivity}
\end{align}
Here, we introduced compact notation: $[A,B]_{nm} = \sum_{l \neq n, m} A_{nm} B_{mn} - (B \lra A)$, and $\Delta^{x,y}_{nm}$ = $v^{x,y}_{nn}-v^{x,y}_{mm}$, which is resolved using the Hadamard product, i.e., $(A\Delta^{x,y})_{nm} = A_{nm} \Delta^{x,y}_{nm}$. $\varepsilon_{n} (\mathbf{k})$ is the energy of the n-th Bloch band, at momentum $\mathbf{k}$, and is also inserted in the expression in the Hadamard form. $\mA^{x,y}$ is the non-Abelian Berry connection, defined as usual $\left \langle n | \hat{\mathbf{r}} | m \right\rangle = \mathcal{A}_{nm}$, where $\mathbf{r}_{nm}$ is the position operator. 

In writing Eq.~\eqref{eq:mainresult}, we split the nonlinear conductivity into three physically distinguishable response types. 
Namely, $I_1$ is associated with a velocity shift, $\Delta_{nm}^\alpha$, while $I_3$ describes a renormalization of the Berry curvature. Each of these is individually gauge invariant (see SI), and is the result of residual processes from the optical, high-frequency limit. The velocity shift $I_1$ has a form similar to the injection current seen in TRS-broken systems at high frequencies \cite{Holder2020,Zhang2019}. $I_3$ is a purely multi-band object seen at higher-order response. In the case of a coupling between magnetic and electric fields, it is related to the non-topological part of the magneto-electric polarizability \cite{Michishita2022}.
Finally, $I_2$ involves a tensor $S^{xy}$ which is related to the shift-vector found in optical response \cite{Kaplan2021a}. This quantity is defined as,
\begin{align}
    S^{\alpha \beta}_{nm} &= (1-\delta_{nm})\left(\lambda^{\alpha\beta}_{nm}-\tfrac{i}{2}\left(\mA^\alpha_{nm} \delta^\beta_{nm} + \mA^\beta_{nm} \delta^\alpha_{nm}\right)\right), \\ 
    \lambda^{\alpha \beta}_{nm} &= \tfrac{i}{2} \left(\left\langle n | \partial_\alpha \partial_\beta m\right\rangle - \left\langle \partial_\alpha \partial_\beta n | m\right\rangle\right).
\end{align}
$\lambda^{\alpha \beta}$ presents a higher derivative on the wavefunction, which results from the resolution of $\partial_\alpha \mA^\beta_{nm}$. $\delta^\alpha_{nm} = \mA_{nn}^\alpha - \mA_{mm}^\alpha$ which encodes a real-space shift of the wave-function center \cite{Morimoto2016} also appears, with the latter entering via a Hadamard product, thus rendering $S^{\alpha \beta}$ manifestly gauge covariant: $S^{\alpha \beta}_{nm} \to e^{i\theta_{nm}(\mathbf{k})}S^{\alpha \beta}_{nm}$, under the ${U(1)}^{N}$ gauge transformation, with the Bloch wavefunctions transforming as $\ket{\psi_{n\mathbf{k}}} \to e^{i\theta_n (\mathbf{k})}\ket{\psi_{n\mathbf{k}}}$. 
The commutator structure ensures the gauge invariance of the entire expression for $\sigma^{xy}$. 
A detailed proof of the gauge invariance under a ${U(1)}^{N}$ transformation for each of the terms is presented in the SI~\cite{Note1}.
The appearance of $\Delta^{x,y}$ as well as $\delta^{x,y}$ in Eq.~\eqref{eq:fconductivity} shows the connection of these objects to expressions at finite frequency such as injection and shift currents \cite{Sipe1999, Young2012, Tan2016, Holder2020}.
The second-order correction in Eq.~\eqref{eq:fconductivity} has several noteworthy properties:

\paragraph{Absence of longitudinal components.---} The correction may be nonzero only in the direction perpendicular to the applied field, and only enters the transverse components of $\sigma^{\alpha \beta}$ (in any dimension). This is of course due to the fact that the correction is related to the Berry curvature, which ensures that the resultant current is always perpendicular to the perturbation. Consequently, the correction does not violate charge conservation nor does it produce a longitudinal response which would require a finite Fermi surface \cite{Holder2021}. \\
\emph{Multiband nature} -- Inspection of Eqs.~\eqref{eq:i1}-\eqref{eq:fconductivity} reveals that the   in-gap conductivity is generated by interband processes, which are due to virtual transitions between occupied and unoccupied bands. This is a direct result of the commutator structure because $f_n [A,B]_{nn} = f_{nm}A_{nm}B_{mn}$ where $f_{nm} = f_n - f_m$. The latter vanishes if both states are occupied or empty. Since $\Omega^z_{nn}$ can be projected into a single-band, it appears at linear order. But corrections to this are manifestly multi-band objects, involving direct probes of the states through $\Delta^\alpha_{nm}, \varepsilon_{nm}, S^{\alpha \beta}_{nm}$. The presence of the shift tensor $S^{\alpha \beta}_{nm}$ suggests a property which is encoded in at least two bands, as the commutator in which it appears restricts $n \neq m$. Furthermore, we note the presence of a higher-order multi-band term, $\left[\mA^x \varepsilon^{-1}, \Omega^z\right]_{nn}$, which is $(I_3)_{nn}$ in Eq.~\eqref{eq:fconductivity}. To parse this object, one evaluates $\Omega^z_{nm}$, where $n \neq m$. Using the definition of the commutator, however, $[A, B]_{nm} = \sum_{l \neq n,m} A_{nl} B_{lm} - (a \lra B)$. From this, it follows that this term only exists for three bands or more. In the two band limit, the commutator can be directly evaluated to be $[A, B]_{12} = \sum_{l \neq 1,2} A_{1l}B_{l2} - (A \leftrightarrow B) = 0$, as the sum cannot extend over \emph{any} intermediate state. This term represents, therefore, a unique signature of a quantum process which involves interband transitions between two principle bands -- occupied and empty -- with an assisting interim third band. 

\paragraph{Symmetries.---} The correction to the anomalous Hall conductivity strongly depends on the underlying symmetry of the crystal lattice. Firstly, a general requirement for the appearance of intrinsic in-gap responses is the breaking of time-reversal symmetry (TRS). Since this effect is quadratic in the electric field, inversion symmetry ($P$) must be broken as well. The symmetry discussion is simplified by considering the correction as a \emph{second-order} Hall conductivity. We define $\sigma^{xx;y} = \frac{\delta^2 j_y}{\delta E_x \delta E_x}$. Eqs.~\eqref{eq:i1}-\eqref{eq:fconductivity} show that the Hall part of the tensor $\sigma^{ab;c}$ takes the form $\sigma^{aa;c}$.
Applying the von Neumann principle \cite{Tinkham2003}, we find that for all rotational symmetries $C_{n,z}, n \geq 2$, $\sigma^{aa;c}$ vanishes identically. In 3D (or higher), other components of the response tensor are permitted, e.g. of the form $\sigma^{xx;z}$. The emergence of a longitudinal in-gap current is restricted by the presence of point-group symmetries. In the case of $C_{3z}$, for example, $\sigma^{xx;y} = - \sigma^{yy;y}$, but since the correction vanishes for $\sigma^{yy;y}$, the Hall response is null as well. 
\paragraph{Twisted bilayer graphene.---}
As a candidate system to test our results, we suggest to use strained twisted bilayer graphene (TBG) \cite{Bistritzer2011,Santos2012, Carr2018,He2020}. As previously seen in the case of resonant optical conductivity \cite{Kaplan2021a}, in TBG second-order electrical responses can become exceptionally large due to the large phase space for transitions between flat bands. We model a time-reversal breaking state of TBG by considering the Bistrizer-MacDonald \cite{Bistritzer2011} continuum model for a single valley and spin of TBG at a twist angle of $\theta = 1.05^{o}$. 
\begin{figure}
    \centering
    \includegraphics[width=1\columnwidth]{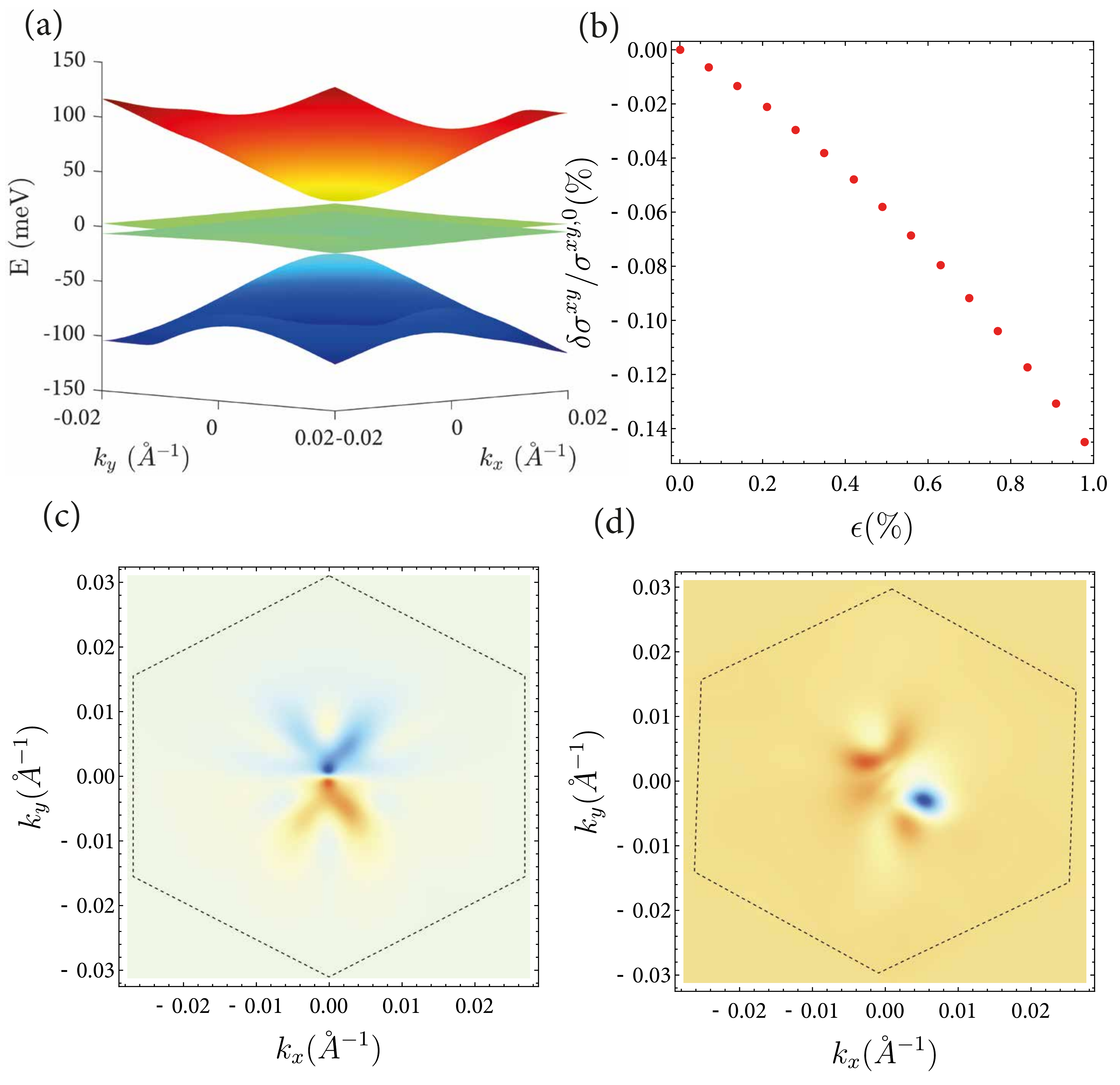}
    \caption{\textbf{Nonlinear correction to the anomalous Hall in TBG.} (a) Band structure of twisted bilayer graphene for $\theta \sim 1.05^{o}$. Remote dispersive bands contribute to the interband correction, which vanishes in the single-band limit. (b) Magnitude of the correction to the AHC $\sigma^{xy,0} = \frac{e^2}{h}$ as a function of applied uniaxial strain $\epsilon$ in TBG. (c-d) Momentum space distribution of $\delta \sigma^{xy}$ for $\epsilon = 0$ (c) and $\epsilon = 0.8\%$ in (d).
    }
    \label{fig:fig2}
\end{figure}
This corresponds to experimental measurements of TBG in the ferromagnetic, 3/4 filled state \cite{Sharpe2019,Serlin2020}, in a series of cascading symmetry-broken states in this system~\cite{Zondiner2020,Liu2021}. This phase is topological with a Chern number $C_N = 1$. In a sample, the TBG is usually placed on top of a layer of hBN \cite{Lee2016,Kim2018}, which breaks inversion symmetry. This can be modelled by introducing a staggered potential $\Delta = 17 \textrm{meV}$~\cite{He2020}. 
Since TBG on top of hBN still retains a $C_{3z}$-symmetry, the correction considered here remains zero. 
This is shown in Fig.~\ref{fig:fig2}(b). Fig.~\ref{fig:fig2}(c) also reveals that the momentum distribution of the correction is anti-symmetric within the mini Brillouin zone (mBZ) and thus vanishes after integration. However, introducing strain breaks $C_{3z}$, rendering the correction in Eq.~\eqref{eq:mainresult} nonzero, as seen in Fig.~\ref{fig:fig2}(d), with the mBZ modified as well. The deviation from $\sigma^{xy}_0 = \frac{e^2}{h}$ increases with increasing strain. Using a typical strain amplitude of $\epsilon \sim 0.65\%$ \cite{Kerelsky2019}, and electric field strengths of $E=300 V m^{-1}$, it reaches a value of $0.1\%$, which is comparable to the deviation from perfect quantization in recent experiments~\cite{Serlin2020}. 

\paragraph{Semiclassical interpretation.---} The structure of the correction 
permits the following semiclassical form. We define the electric field-induced shift tensors,
\begin{align}
    \mathsf{v}_E^a = e\frac{\mathcal{A}^a \bm{\Delta}^b}{\varepsilon} E_b, ~\mathsf{S}_E^a  = e\bm{S^{ab}} E_b, ~\mathsf{\Omega}_E = e\Omega^{ab}E_b,
\end{align}
Where all terms enter as Hadamard products. In band basis, these objects are translated as $\mathsf{v}^a_E = \sum_b \frac{e}{\varepsilon_{nm}} \mathcal{A}^{a}_{nm} \Delta^b_{nm} E_b$. This can be carried out analogously for all terms. The semi-classical anomalous current at second order can then be written as
\begin{align}
    \mathbf{j} = \frac{e^2}{\hbar}\sum_{n \in \textrm{occ.}} \mathbf{E} \times \int \frac{\textrm{d}^2 k}{(2\pi)^2} \left(\mathbf{\Omega}_{nn} + \left[\frac{\mathcal{A}}{\varepsilon} \times  \mathsf{\bm{V}}\right]_{nn}\right).
    \label{eq:semiclass}
\end{align}
Here, $\displaystyle \mathsf{\bm{V}}^a_{E} = \mathsf{v}_E^a + \mathsf{S}_E^a + \mathsf{\Omega}_E^a$. The cross product is to be interpreted as usual \cite{Shi2007,Xiao2010}, such that $\left[\frac{\mathcal{A}}{\varepsilon}\times\mathsf{V}\right]_{nn} = \sum_{m \in \textrm{unocc.}} \epsilon^{abc}\frac{\mathcal{A}^b_{nm}}{\varepsilon_{nm}} \mathsf{V}^c_{mn,E} - (n \lra m)$. Here $\epsilon^{abc}$ is the Levi-Civita symbol. 
This partition into three pieces is identical in content to the previous decomposition into $I_{1}$, $I_{2}$, $I_{3}$ in Eq.~\eqref{eq:mainresult}.
$\frac{\mathsf{\bm{V}}^a}{\hbar}$ carries units of velocity, meaning that it is the velocity of the (instantaneous) charge displacement upon application of the external electric field $\mathbf{E}$. 
At first order, this displacement modifies the position operator through a change in the charge dipole. 
At second order in the applied field, this deformation couples back to the position operator, resulting in a correction to the anomalous velocity, now effectively quadratic in the applied field. 
The weight $\varepsilon_{nm}^{-1}$ attached to the position operator reflects the quantum-perturbative expansion, since the $\mathsf{V}^a$ is now explicitly \textit{inter-band}, and the transitions to the neighboring bands are suppressed by the energy gap. 
The single band limit can recovered in the limit where all unoccupied bands are infinitely separated from the top of the valence band 
such that $\varepsilon_{nm} \to \infty$. 
In this limit $\frac{\mathcal{A}}{\varepsilon}$ vanishes, and the usual anomalous velocity is recovered. 
A visualization of the momentum-space structure of $\Re{\mathsf{V}}$ using a simplified two band model can be found in the SI~\cite{Note1}. 
\paragraph{Robustness of the integer quantum Hall effect.---} The precise quantization of the conductivity $\sigma^{xy}$ of the integer quantum hall effect for a 2D electron gas can be understood in the absence of any higher-order corrections at finite bias. To show that our correction vanishes identically for Landau levels, consider the Hamiltonian of an electron gas in the Landau gauge, $\mathbf{A} = (0, -Bx,0)$,
\allowdisplaybreaks
\begin{align}
    H = \frac{p^2_{x'}}{2M} + \frac{M\omega_c^2 x'^2}{2}.
    \label{eq:landauham}
\end{align}
Here as usual $\omega_c = \frac{eB}{M}$, and $x' = x + \frac{\hbar k_y}{eB}$.
We shall show that the quantization of the Hall conductivity is guaranteed by the ladder operator structure. The velocity matrix elements in the Landau level basis are,
\begin{align}
    & v_x = \frac{\partial H}{\partial \pi_{x'}} = \frac{p_{x'}}{M} = i\sqrt{
    \frac{\hbar\omega_c}{2M}}(a^\dagger-a), \\
    & v_y = \frac{\partial H}{\partial \pi_y} = \omega_c x' = \frac{\omega_c l_B}{\sqrt{2}} (a + a^\dagger).
\end{align}
We define the latter operators $a = \frac{1}{\sqrt{2m\hbar\omega_c}}\left(i p_{x'} + M\omega_c x' \right)$ and $l_B = \sqrt{\frac{\hbar}{eB}}$. For Landau levels, $\varepsilon_n = \hbar \omega_c(n+1/2)$, $a^\dagger \ket{n} = \sqrt{n+1}\ket{n+1}$, and $a \ket{n} = \sqrt{n} \ket{n-1}$. The expectation values become $\braket{n}{v_{x}}{m} = i\frac{\hbar}{\sqrt{2}M l_B} \left(\sqrt{m+1}\delta_{n,m+1}-\sqrt{m}\delta_{n,m-1}\right)$, $\braket{n}{v_{y}}{m} = \frac{\omega_c l_b}{\sqrt{2}}\left(\sqrt{m+1}\delta_{n,m+1}+\sqrt{m}\delta_{n,m-1}\right)$. 
The quantization of the linear conductivity is directly related to the ladder structure of operator algebra in the integer quantum Hall fluid. A demonstration of this property is relegated to the SM. However, the fact that the ladder operators only connect Landau levels with energy differences $\Delta  \varepsilon = \pm \hbar \omega_c$ can be used to show that \emph{all} higher-order corrections vanish for the 2D electron gas at high magnetic field. At ~\nd order, the relevant diagrams of the quantum perturbative calculation give two contributions at order $\tau^{0}$ (all other terms vanish in the gapped phase identically) \cite{Kaplan2022a}. For the Hall response tensor $\sigma^{xx;y}$,
\begin{align}
    \notag \sigma^{xx;y} &= \frac{e^3}{\hbar^2} \sum_{n} f_n\left[-2 \left[\varepsilon^{-3} w^{xx}, v_y\right]_{nn} -\left[\varepsilon^{-3} v^x,w^{xy}\right]_{nn} \right] \\ \notag & +  \frac{e^3}{\hbar^2}\sum_{n,m} \biggl[ -4 \frac{f_{nm}}{\varepsilon_{nm} \varepsilon_{nl}^3}v^x_{nm}v^x_{ml}v^y_{ln}  \\ & 
    - 2 \frac{f_{nm}}{\varepsilon_{nm}^2 \varepsilon_{nl}^2}v^x_{nm}v^x_{ml}v^y_{ln} 
    -\frac{f_{nm}}{\varepsilon_{nm}^3 \varepsilon_{nl}}v^x_{nm}v^x_{ml}v^y_{ln}\biggr].
    \label{eq:full2ndorder}
\end{align}
The elimination of the first two commutators in Eq.~\eqref{eq:full2ndorder} is due to the free fermion dispersion of the Landau levels giving $w^{ab}_{nm} = \braket{n}{\frac{\partial^2 H}{\partial {\pi_a} \partial_{\pi_b}}}{m} \propto \delta_{nm}$ since the underlying dispersion is quadratic in Eq.~\eqref{eq:landauham}. The commutator $[A,B]_{nn}$ contains only off-diagonal components of $A_{nm}, B_{nm}$. 
Consequently, since $w^{ab}_{nm} = 0$ for any $a,b, n \neq m$ this contribution vanishes. We are left with the triple product $v^x_{nm} v^x_{ml} v^y_{ln}$. By applying the ladder structure for $v^{x,y}_{nm}$ the following combinations appears: $\delta_{n,m \pm 1}\delta_{m, l \pm 1}\delta_{l,n \pm 1}$~\cite{a_Kaplan2021}. 
By applying, e.g., the middle Kronecker delta, we have the condition that $n = l\pm 1 \pm 1$, and $n = l\mp 1$. Clearly, there exists no $l, n$ that satisfies this constraint. 
This results in $\sigma^{xx;y} = 0$ regardless of the exact structure of the Hamiltonian, provided the algebra of the ladder operators is preserved. 
The generalization of the above can be made by considering that $n$-th order response will contain an $n+1$ product of velocity operators $v^x_{m_1 m_2}v^x_{m_2 m_3}v^x_{m_3 m_4}\ldots v^y_{m_{n}m_{1}}$, which produces the condition that $\delta_{n_1 n_2 \pm 1}\delta_{n_2 n_3 \pm 1}\delta_{n_3 n_4 \pm 1} \ldots$ which yields zero for the real part of the current at any order. 
The only nonzero combination for which band indices can be selected appears at order $n=1$ corresponding to linear response, which gives the quantized integer Hall conductivity.

\paragraph{Conclusions.---} We have shown that in general magnetic insulators which break inversion, time reversal as well as rotational symmetries, a quadratic correction to the in-gap Hall conductivity appears. In a topological phase, this indicates that measurements at finite bias will deviate from the quantized value due to the presence of nonlinear corrections. As an example, we calculated the correction for strained twisted bilayer graphene, 
finding for the magnitude of the nonlinearity values which are comparable with the observed precision of the quantization in the recent experiment of Ref.~\cite{Serlin2020}.
Another experimental signature may appear in the non-reciprocal nature of the conductivity. Namely, in systems where the correction is observable we find that $\sigma^{xy} \neq - \sigma^{yx}$, and the sum $\sigma^{xy} + \sigma^{yx}$ can thus be treated as a proxy for the correction. Thirdly, the quantities derived here might be visible as non-linear powers in the $I-V$ curve. 

Recent progress on nonlinearities in graphene superlattices \cite{Berdyugin2022} suggests that experiments at moderate finite bias on graphene-based systems are possible. By tuning the graphene superlattices to the QAH state and sweeping the bias, the nonlinear corrections, as well as the non-reciprocity they produce might be accessible. In addition, the sensitivity of the effect to strain suggests an electro-mechanical setup in which a controlled application of tensile stress is employed in order to modify the Hall conductivity (at finite bias). We note that systems with $C_{3z}$ symmetry, such as doped Bi\textsubscript{2}Se\textsubscript{3} \cite{Chang2013} do not exhibit this correction due to the symmetry restriction. 
Our results might be relevant in understanding why experiments on systems with rotational symmetries observe a much more precisely quantized QAHE~\cite{Chang2015high,Okazaki2022}.
Related to that, the reasoning presented here raises the question whether third or even higher order corrections are non-vanishing even if a QAH system has inversion and $C_3$ symmetry.  Our result establishes a concrete difference in the quantization of the QAHE compared to the IQHE, which suggests that using QAHE systems for metrology depends on subtleties related to the crystal systems, symmetries and the magnitude of the applied bias. Our results predict a striking phenomenon that a generic insulator can present a nonlinear current response in the dc limit.


\begin{acknowledgements}
We thank
Ady Stern and Xi Dai
for useful discussions.
B.Y.\ acknowledges the financial support by the European Research Council (ERC Consolidator Grant No. 815869, ``NonlinearTopo'') and Israel Science Foundation (ISF No. 2932/21). 
D. Kaplan appreciates support from the Weizmann Institute Sustainability and Energy Research Initiative. \end{acknowledgements}

%

\par
\newpage
\pagebreak
\newpage
\onecolumngrid
\begin{center}
    \textbf{\large Supplementary information}
\end{center}
\setcounter{equation}{0}
\setcounter{figure}{0}
\setcounter{table}{0}
\setcounter{page}{1}
\makeatletter
\renewcommand{\theequation}{S\arabic{equation}}
\renewcommand{\thefigure}{S\arabic{figure}}
\renewcommand{\bibnumfmt}[1]{[S#1]}
\section{Reduction of the Kubo formula}
In this section we give the derivation of the electric-field induced nonlinear correction via a Kubo formula. The details of the diagrammatic approach are listed in Refs. \cite{Holder2020}. We specifically the insertion of finite lifetimes according to the prescription of Ref. \cite{Kaplan2022a}, and employ the notation therein. We expand to order $\tau^{0}$, which represents the dissipation-less correction to the linear conductivity. In the diagrammatic picture, 4 diagrams contribute (may be insert these later). For simplicity, we consider here the case of $\sigma^{xx;y}$. An analogous expression can be derived for $\sigma^{yy;x}$. For compactness, we omit the Fermi occupation factor throughout. The total conductivity $\sigma^{xx;y}$ at order $\tau^{0}$ reads,
\begin{align}
    \sigma^{xx;y} = \mathcal{W}^{xx;y}  + \mathcal{V}^{xx;y}.
\end{align}
$\mathcal{W}^{xx;y}$ contains contributions from two-photon vertices, while $\mathcal{V}^{xx;y}$ are three-legged diagrams. Parsing $\mathcal{W}^{xx;y}$,
\begin{align}
    \mathcal{W}^{xx;y}_{\tau^{0}} = -2 \left[\varepsilon^{-3} w^{xx}, v_y\right] - \left[\varepsilon^{-3} v^x,w^{xy}\right] = -2 \left[\varepsilon^{-3} v_y, w^{xx}\right] - \left[\varepsilon^{-3} v^x,w^{xy}\right].
\end{align}
Recall that $w^{xx} = 2i \Delta^x \mA^x - \left[\varepsilon \mA^x, \mA^x\right] + i\varepsilon S^{xx}$. Furthermore, $w^{xy} = i \Delta^x \mA^y + i \Delta^y \mA^x - \frac{1}{2}\left[\varepsilon \mA^x, \mA^y\right] - \frac{1}{2} \left[\varepsilon \mA^y, \mA^x\right] + i\varepsilon S^{xy}$. $v^{x,y} = i\varepsilon \mA^{x,y}$ (since only off-diagonal components are involved). 
\begin{align}
   \notag \mathcal{W}^{xx;y}_{\tau^{0}} &= -2 i \left[\varepsilon^{-2} \mA^y, 2i \Delta^x \mA^x - [\varepsilon \mA^x, \mA^x] + i\varepsilon S^{xx}\right] - \\ & 
    \quad i\left[\varepsilon^{-2} \mA^x, i \Delta^x \mA^y +  i \Delta^y \mA^x -(1/2)[\varepsilon \mA^x, \mA^y] -(1/2)[\varepsilon \mA^y, \mA^x] + i\varepsilon S^{xy}\right].
\end{align}
Terms may now be rearranged given the transposition properties of objects inside the commutator. For example, $\left[\varepsilon^{-2} \mA^x, \Delta^x \mA^y\right] = \left[\varepsilon^{-2} \mA^{y}, \Delta^x \mA^x\right]$. This stems from the fact that $\varepsilon_{nm}^2 = \varepsilon_{mn}^2$, but $\Delta^x_{nm} = -\Delta^x_{mn}$. Thus,
\begin{align}
  \notag  \mathcal{W}^{xx;y}_{\tau^{0}} &= 5 \left[\varepsilon^{-2} \mA^y, \Delta^x \mA^x\right] + \left[\varepsilon^{-2} \mA^x, \Delta^y \mA^x\right] + 2i \left[\varepsilon^{-2} \mA^y, \left[\varepsilon \mA^x, \mA^x\right]\right] - 2 \left[\varepsilon^{-1} \mA^y, S^{xx}\right] + \\ &
  \frac{i}{2} \left[\varepsilon^{-2} \mA^x, \left[\varepsilon \mA^x, \mA^y\right] + \left[\varepsilon \mA^y, \mA^x\right]\right] - \left[\varepsilon^{-1} \mA^x, S^{xy}\right].
\end{align}
Next we turn our to $\mathcal{V}^{xx;y}$. As the expressions contain denominators $\varepsilon_{nl}$ which depend on an intermediate index, we first isolate two cases of interest: $l = n$, $l = m$. The remainder are pieces for which $l \neq n,m$ and therefore are amenable to being written as proper commutators, as noted in the introduction. Since this section will involve explicit diagonal parts of the velocity operators $v^{x,y}_{nn}$ we restore the Fermi occupation factors.
Firstly,
\begin{align}
    \mathcal{V}^{xx;y}_{l=n,\tau^0} = \frac{f_{nm}}{2\varepsilon_{nm}^4} v^x_{nm} v^x_{mn} v^y_{nn} + \frac{f_{nm}}{2\varepsilon_{nm}^4} v^y_{nn}v^x_{mn}v^x_{nm}
\end{align}
Interchanging the summation on the second term $(n \lra m)$, gives,
\begin{align}
\frac{f_{nm}}{2\varepsilon_{nm}^4} \left( v^x_{nm} v^x_{mn} (v^y_{nn} - v^y_{mm})\right) = \frac{1}{2} \left[\frac{v^x}{\varepsilon^4} \Delta^y, v^x\right] = \frac{1}{2}\left[\frac{\mA^x}{\varepsilon^2} \Delta^y, \mA^x\right].
\end{align}
Next is the case of $l = m$,
\begin{align}
    \mathcal{V}^{xx;y}_{l=m,\tau^0} = -7 \frac{f_{nm}}{\varepsilon_{nm}^4} v^x_{nm} v^x_{mm} v^y_{mn} - 7\frac{f_{nm}}{\varepsilon_{nm}^4} v^y_{nm}v^x_{mm}v^x_{mn} = 7 \left[\varepsilon^{-4} v^y\Delta^x,v^x\right] = 7 \left[\varepsilon^{-2} \mA^y\Delta^x,\mA^x\right] = -7 \left[\varepsilon^{-2} \mA^y,\Delta^x\mA^x\right].
\end{align}
The remaining terms, such that $l \neq n,m$ are,
\begin{align}
   \mathcal{V}^{xx;y, (1)}_{l \neq n,m,\tau^0} =  -4 \frac{f_{nm}}{\varepsilon_{nm} \varepsilon_{nl}^3}v^x_{nm}v^x_{ml}v^y_{ln} - 2 \frac{f_{nm}}{\varepsilon_{nm}^2 \varepsilon_{nl}^2}v^x_{nm}v^x_{ml}v^y_{ln} - \frac{f_{nm}}{\varepsilon_{nm}^3 \varepsilon_{nl}}v^x_{nm}v^x_{ml}v^y_{ln},
   \label{eq:supp_totalv}
\end{align}
while $\mathcal{V}^{xx;y, (2)}_{l \neq n,m}$ is the complex conjugate of $\mathcal{V}^{xx;y, (1)}_{l \neq n,m}$. Let us treat each term separately. After substituting $v^{x,y}$, and adding the complex conjugate the first term yields $-4 \frac{f_{nm}}{\varepsilon_{nm} \varepsilon_{nl}^3}v^x_{nm}v^x_{ml}v^y_{ln} = -4i \left[\mA^x, \left[\varepsilon\mA^x, \varepsilon^{-2} \mA^y\right]\right]$. We break this term up into two pieces of equal prefactor, 2, and the first piece is replaced using a Jacobi identity. That is, 
$-2i \left[\mA^x, \left[\varepsilon\mA^x, \varepsilon^{-2} \mA^y\right]\right] = 2i \left[\varepsilon^{-2} \mA^y, \left[\mA^x, \varepsilon \mA^x\right]\right] +2i \left[\varepsilon \mA^x, \left[\varepsilon^{-2} \mA^y, \mA^x\right]\right]$. The other piece is written down differently. Using the fact that $\varepsilon_{ml} = \varepsilon_{mn} + \varepsilon_{nl}$, $-2i \left[\mA^x, \left[\varepsilon\mA^x, \varepsilon^{-2} \mA^y\right]\right] = -2i\left[\mA^x, \varepsilon\left[\mA^x, \varepsilon^{-2} \mA^y\right]\right]+2i\left[\mA^x, \left[\mA^x, \varepsilon^{-1} \mA^y\right]\right]$. Adding this up once more gives,
\begin{align}
    -4i \left[\mA^x, \left[\varepsilon\mA^x, \varepsilon^{-2} \mA^y\right]\right] = 2i \left[\varepsilon^{-2} \mA^y, \left[\mA^x, \varepsilon \mA^x\right]\right] + 2i\left[\mA^x, \left[\mA^x, \varepsilon^{-1} \mA^y\right]\right].
\end{align}
The next term in Eq.~\eqref{eq:supp_totalv} is decomposed analogously. 
\begin{align}
-2\frac{f_{nm}}{\varepsilon_{nm}^2 \varepsilon_{ln}^2}v^x_{nm}v^x_{ml}v^y_{ln} = 2i \left[\frac{\mA^x}{\varepsilon}, \left[\varepsilon \mA^x, \varepsilon^{-1}\mA^y\right]\right] = -2i \left[\mA^x,\left[\mA^x, \varepsilon^{-1}\mA^y\right]\right] - 2i\left[\varepsilon^{-1} \mA^x,\left[\mA^x, \mA^y\right]\right]
\end{align}
The last term in Eq.~\eqref{eq:supp_totalv} is given by,
\begin{align}
    -\frac{f_{nm}}{\varepsilon_{nm}^3 \varepsilon_{nl}}v^x_{nm}v^x_{ml}v^y_{ln} = -i\left[\varepsilon^{-2} \mA^x, \left[\varepsilon \mA^x, \mA^y\right] \right].
\end{align}
We are now ready to assemble all the pieces we have. We combine,
\begin{align}
     &\notag \mathcal{V}^{xx;y} + \mathcal{W}^{xx;y} = \\ 
     & \notag 5 \left[\varepsilon^{-2} \mA^y, \Delta^x \mA^x\right] + \left[\varepsilon^{-2} \mA^x, \Delta^y \mA^x\right] + 2i \left[\varepsilon^{-2} \mA^y, \left[\varepsilon \mA^x, \mA^x\right]\right] - 2 \left[\varepsilon^{-1} \mA^y, S^{xx}\right] + \\ \notag  &
  \frac{i}{2} \left[\varepsilon^{-2} \mA^x, \left[\varepsilon \mA^x, \mA^y\right] + \left[\varepsilon \mA^y, \mA^x\right]\right] - \left[\varepsilon^{-1} \mA^x, S^{xy}\right] + \frac{1}{2}\left[\frac{\mA^x}{\varepsilon^2} \Delta^y, \mA^x\right] -7 \left[\varepsilon^{-2} \mA^y,\Delta^x\mA^x\right] + 2i \left[\varepsilon^{-2} \mA^y, \left[\mA^x, \varepsilon \mA^x\right]\right]\\ & 
   + 2i\left[\mA^x, \left[\mA^x, \varepsilon^{-1} \mA^y\right]\right] -2i \left[\mA^x,\left[\mA^x, \varepsilon^{-1}\mA^y\right]\right] - 2i\left[\varepsilon^{-1} \mA^x,\left[\mA^x, \mA^y\right]\right] -i\left[\varepsilon^{-2} \mA^x, \left[\varepsilon \mA^x, \mA^y\right] \right].
\end{align}
We obtain,
\begin{align}
     &\notag \mathcal{V}^{xx;y} + \mathcal{W}^{xx;y} = \\ 
     & \notag -2\left[\varepsilon^{-2} \mA^y, \Delta^x \mA^x\right] + \frac{1}{2}\left[\varepsilon^{-2} \mA^x, \Delta^y \mA^x\right] - 2 \left[\varepsilon^{-1} \mA^y, S^{xx}\right] + \frac{i}{2} \left[\varepsilon^{-2} \mA^x, -\left[\varepsilon \mA^x, \mA^y\right] + \left[\varepsilon \mA^y, \mA^x\right]\right] + \\ 
     & - \left[\varepsilon^{-1} \mA^x, S^{xy}\right] - 2i\left[\varepsilon^{-1} \mA^x,\left[\mA^x, \mA^y\right]\right].
\end{align}
This term contains several total derivatives, which are Fermi surface terms, and give rise to the recently proposed gravitational anomaly and intrinsic non-dissipative Hall effects \cite{Holder2021,Gao2014}. We remove them by observing that,
\begin{align}
    \partial_y \left[\varepsilon^{-1} \mA^x, \mA^x\right] = 2 \left[\varepsilon^{-1} \mA^x, S^{xy}\right]+\left[\varepsilon^{-2} \mA^x, \Delta^y \mA^x\right] +\left[\varepsilon^{-1} \mA^x, i [\mA^y, \mA^x
    ]\right] \\
    \partial_x \left[\varepsilon^{-1} \mA^y, \mA^x\right] =  \left[\varepsilon^{-1} \mA^x, S^{xy}\right] + \left[\varepsilon^{-1} \mA^y, S^{xx}\right] -\left[\varepsilon^{-2} \mA^y \Delta^x, \mA^x\right] + \left[\varepsilon^{-1} \mA^x, (i/2) [\mA^x, \mA^y]\right].
\end{align}
By combining these identities with the symmetrization condition one finds,
\begin{align}
    \notag \sigma^{xx;y} &= 2 \partial_y \left[\varepsilon^{-1} \mA^x, \mA^x\right] -
    \partial_x \left[\varepsilon^{-1} \mA^y, \mA^x\right] + \\ &  \frac{1}{2} \left[\varepsilon^{-2} \mA^y, \Delta^x \mA^x\right] - \frac{1}{2}\left[\varepsilon^{-2} \mA^x, \Delta^y \mA^x\right] + \left[\varepsilon^{-1} \mA^x, S^{xy}\right] - \left[\varepsilon^{-1} \mA^y, S^{xx}\right] + \left[\varepsilon^{-1} \mA^x, (i/2) [\mA^x, \mA^y]\right].
\end{align}
In the above, we employed the identity that $\tilde{\Omega}^{xy,1} - \tilde{\Omega}^{yx,1} = \varepsilon \Omega^{ab}$, otherwise proven here \cite{Kaplan2022a}. We note that this form is fully compatible with the consistent separation of nonlinear response conductivity into Hall components carried out in Ref. \cite{Tsirkin2021}. We also note that the positional shift $S^{xy}$ described here is related (while more general) to the quantum metric dipole, also recently shown to present an in-gap Hall conductivity \cite{Lahiri2022}.
\section{Gauge invariance of the derived correction}
The introduction of derivatives of the wavefunction in the expressions for $I_1-I_3$
Eqs.~\eqref{eq:i1}-\eqref{eq:fconductivity} requires delicate handling of gauge transformations. The Bloch manifold of cell-periodic states is characterized by an invariance to the $U(1)^N$ transformation,
\begin{align}
    \ket{n\mathbf{k}} \to e^{-i\theta_n(k)} \ket{n\mathbf{k}}.
\end{align}
For notational ease, we suppress below the label $\mathbf{k}$, and refer to $\ket{n}$, as $\ket{n} = \ket{n \mathbf{k}}$. 
The Bloch periodic part of the Hamiltonian commutes with this gauge transformation since $[H(\mathbf{k}), f(k)] = 0$. The observables of optical response are modified due to the gauge covariance of the states. The velocity operator $v^\alpha \to U v^\alpha$, where $U_{nm} = e^{i \theta_{nm}}$, $\theta_{nm} = \theta_n - \theta_m$. Clearly, only diagonal components, such as those comprising $\Delta^\alpha_{nm} = v^\alpha_{nn} - v^\alpha_{mm}$, are automatically gauge invariant. Products such as $(v^\alpha)^\dagger v^\beta \to v^\alpha U^\dagger U v^\beta  = v^\alpha v^\beta$ are gauge invariant. In this respect, the equations of optical response in the velocity gauge are manifestly gauge invariant, as they combine products of gauge covariant operators derived from the Hamiltonian. The issue of local $U(1)$ gauge invariance in the context of electromagnetism and optical response has recently been addressed, with gauge invariance formally proven \cite{Ventura2017, Passos2018,Joao2020,Rostami2021}. Our primary focus, therefore, is to show that Eqs.~\eqref{eq:i1}-\eqref{eq:fconductivity} which involve \emph{derivatives} of the Bloch periodic part of the electronic wavefunctions are also gauge invariant. The principle derivatives defining 2nd order optical response are \cite{Kaplan2022a}:
\begin{align}
    \mathcal{A}^\alpha_{nm} &= \braket{n}{i\partial_\alpha}{m} \to  \partial_\alpha \theta_n \delta_{nm} + e^{i\theta_{nm}}
    \label{eq:berry_trans}\braket{n}{i\partial_\alpha}{m}  = e^{i\theta_{nm}}\mathcal{A}^\alpha_{nm} +  \partial_\alpha \theta_n \delta_{nm} \\ 
    \lambda^{\alpha \beta}_{nm} &= \frac{1}{2}\braket{n}{i\partial_\alpha \partial_\beta}{m} + (\textrm{c.c.}, n \lra m) \to e^{i\theta_{nm}}\lambda^{\alpha \beta}_{nm} + \frac{ie^{i\theta_{nm}}}{2}\left(\mathcal{A}^\alpha_{nm} \partial_\beta \theta_{nm}+\mathcal{A}^\beta_{nm} \partial_\alpha \theta_{nm}\right), ~ n\neq m.
\end{align}
We define objects which are gauge \emph{covariant} as those which transform as $A_{nm} \to A_{nm}e^{i\theta_{nm}}$. Consequently, combinations of the form $A_{nm}A_{mn}$ are manifestly gauge \emph{invariant} since they transform like the velocity operator, as shown above, or $e^{i\theta_{nm}}A_{nm} e^{i\theta_{mn}}A_{mn} = A_{nm}A_{mn}$. Generally, for \emph{any} two covariant objects $A, B$, the commutator of the two satisfies,
\begin{align}
    [A,B]_{nm} \to e^{i\theta_{nm}}[A,B]_{nm},
\end{align}
rendering its diagonal part gauge invariant. This follows from the definition introduced in the main text, $[A,B]_{nm} = \sum_{l \neq n,m} A_{nl}B_{lm} - (A \lra B)$. We now prove that the quantity $S^{\alpha \beta}_{nm}, n \neq m$ is gauge covariant, which appears in $I_2$, of Eq.~\eqref{eq:i2}. We note that $S^{\alpha \beta}$ consists of two portions: $\lambda^{\alpha \beta}_{nm}$ and the Hadamard product, $\frac{i}{2}\mathcal{A}^\alpha_{nm} \delta^\beta_{nm} + (\alpha \lra \beta)$. Tackling the latter first,
\begin{align}
    \frac{i}{2}\mathcal{A}^\alpha_{nm} \delta^\beta_{nm} \to \frac{i}{2} e^{i\theta_{nm}} \mathcal{A}^\alpha_{nm} \partial_\beta \theta_{nm}.
\end{align}
To uncover the transformation properties of $\lambda^{\alpha \beta}_{nm}$ for $n \neq m$, we first observe that under $\ket{m} \to e^{i\theta_m} \ket{m}$, $\partial_\alpha \partial_\beta \ket{m} \to e^{i\theta_{m}}\left(i \partial_\alpha \theta_m \ket{\partial_\beta m} +i \partial_\beta \theta_m \ket{\partial_\alpha m} + \ket{\partial_\alpha \partial_\beta m}\right)$. We explicitly remove the term $e^{i\theta_{m}}\partial_\alpha \partial_\beta \theta_m \ket{m}$ since we assume that $n \neq m$ and this contribution must vanish when projected back onto the Bloch states. Multiplying on the left with $i\bra{n}$, we find that,
\begin{align}
    \lambda^{\alpha \beta}_{nm} \to \frac{e^{i\theta_{nm}}}{2} \lambda^{\alpha \beta}_{nm} + \frac{i}{2} e^{i\theta_{nm}} \mathcal{A}^\alpha \partial_\beta \theta_{mn} + (\alpha \lra \beta).
\end{align}
It follows that the sum of the objects,
\begin{align}
    \frac{1}{2}\lambda^{\alpha \beta}_{nm} + \frac{i}{2}A^\alpha_{nm} \delta^\beta_{nm} + (a \lra b) \to e^{i\theta_{nm}} ( \frac{1}{2}\lambda^{\alpha \beta}_{nm} + \frac{i}{2}A^\alpha_{nm} \delta^\beta_{nm}) + (a \lra b),
\end{align}
With the $\theta$ dependent part explicitly cancelling as it appears with opposite indices $\theta_{nm}$ vs $\theta_{mn}$. Lastly, the triple commutator introduced in Eq.~\eqref{eq:fconductivity} transforms using the rules outlined above for commutators. The product reads,
\begin{align}
    \left[\mA^{x}, i[\mA^x, \mA^y]\right]_{nn} =  i\mA^{x}_{nm} \left[\mA^x, \mA^y\right]_{mn} - i \left[\mA^x, \mA^y\right]_{nm}\mA^{x}_{mn} \to e^{i \theta_{nm}} \mA^{x}_{nm} e^{i\theta_{mn}}[\mA^x, \mA^y]_{mn} - (\textrm{c.c.})= \left[\mA^{x}, i[\mA^x, \mA^y]\right]_{nn}.
\end{align}
Here, we used the fact that commutator as defined does not sum over any diagonal contributions and off-diagonal parts of the Berry connection transform according to Eq.~\eqref{eq:berry_trans}.
\section{Quantization of the linear Kubo formula for the IQHE}
The linear response contribution at zero frequency reads \cite{Bernevig2013},
\begin{align}
    \sigma^{xy} = \frac{ie^2}{L_x L_y \hbar} \sum_{n,m}f_{nm} \frac{v^x_{nm} v^y_{mn}}{\varepsilon_{nm}^2}.
    \label{eq:linear_cond}
\end{align}
Here the sum $n,m$ runs over all bands (occupied and empty) including the degenerate manifold of each occupied Landau level. Based on the identities defined in the main text for the velocity matrix elements, we have, 
\begin{align}
    v^x_{nm} v^y_{mn} = \frac{i\hbar \omega_c}{2M} \left(\sqrt{m+1} \delta_{n,m+1}-\sqrt{m}\delta_{n,m-1}\right) \left(\sqrt{m+1} \delta_{n,m+1}+\sqrt{m}\delta_{n,m-1}\right).
\end{align}
The energy difference (measured in frequency units) is $\varepsilon_{nm} = \omega_c (n-m)$. The only surviving terms above are products of equal delta functions. Thus,
\begin{align}
     \frac{v^x_{nm} v^y_{mn}}{\varepsilon_{nm}^2} = \frac{i\hbar \omega_c}{2M} \frac{\left((m+1) \delta_{n,m+1} - m\delta_{n,m-1}\right)}{\omega_c^2 (n-m)^2}.
\end{align}
Assume now that there are $\nu$ occupied bands, which are each $N = \frac{eB L_x L_y}{h}$-fold degenerate. The linear conductivity Eq.~\eqref{eq:linear_cond} has a global Fermi occupation factor difference $f_{nm} = f_n -f_m$. In the bulk of the quantum Hall fluid, a fully flat Landau band is either completely occupied or completely empty. Therefore, $f_{nm}$ is non-zero only in the cases $n = \nu, m=\nu+1, f_{nm} = 1$, and $n = \nu+1, m=\nu, f_{nm} = -1$. By further accounting for the degeneracy, we have,
\begin{align}
\notag
    \sigma^{xy} &= \frac{ie^2}{L_x L_y \hbar} \sum_{n,m}f_{nm} \frac{v^x_{nm} v^y_{mn}}{\varepsilon_{nm}^2} = \\ & \frac{eB L_x L_y}{h} \times \left[\frac{ie^2}{L_x L_y \hbar} \frac{i \hbar \omega_c}{2M} \left(\frac{-(\nu + 1)}{\omega_c^2} - \frac{(\nu + 1)}{\omega_c^2}\right)\right] = \frac{e^2}{h}(\nu + 1).
\end{align}
$\sigma^{xy}$ is therefore quantized by the number of filled Landau levels. 
\section{Two band models}
The simplest approximation that can be made consists of a two band topological model with non-vanishing Berry curvature. We show here that a fundamental condition for the emergence of our derived correction is the presence of nonlinear terms. A minimal Hamiltonian for the which produces a non-zero correction reads,
\begin{align}
  H &= (M+2-\cos(k_x)-\cos(k_y)) \sigma_z ~ +\left(\sin(k_x) + \alpha L_x\right) \sigma_x   +  \left(\sin(k_y) +\beta L_y\right) \sigma_y \\
  &L_x = \sin(k_x)\sin(k_y), ~ L_y = \sin(k_x)\cos(k_y),
  \label{eq:model_ham}
\end{align}
which is a model for a single Dirac cone with nonlinear terms represented by $L_x, L_y$. Here, $\sigma$ are Pauli matrices in an orbital basis. When $\alpha = \beta = 0$, the model retains inversion symmetry, which is defined by $P = \sigma_z (k \to -k)$. This remains true when $\alpha = 0$ alone, since $L_y$ is odd under $(k \to -k)$. The model is topological for all $-2 < M < 0$ with Chern number $C_N = -1$, and is trivial otherwise.
\begin{figure}
    \centering
    \includegraphics[width=0.7\columnwidth]{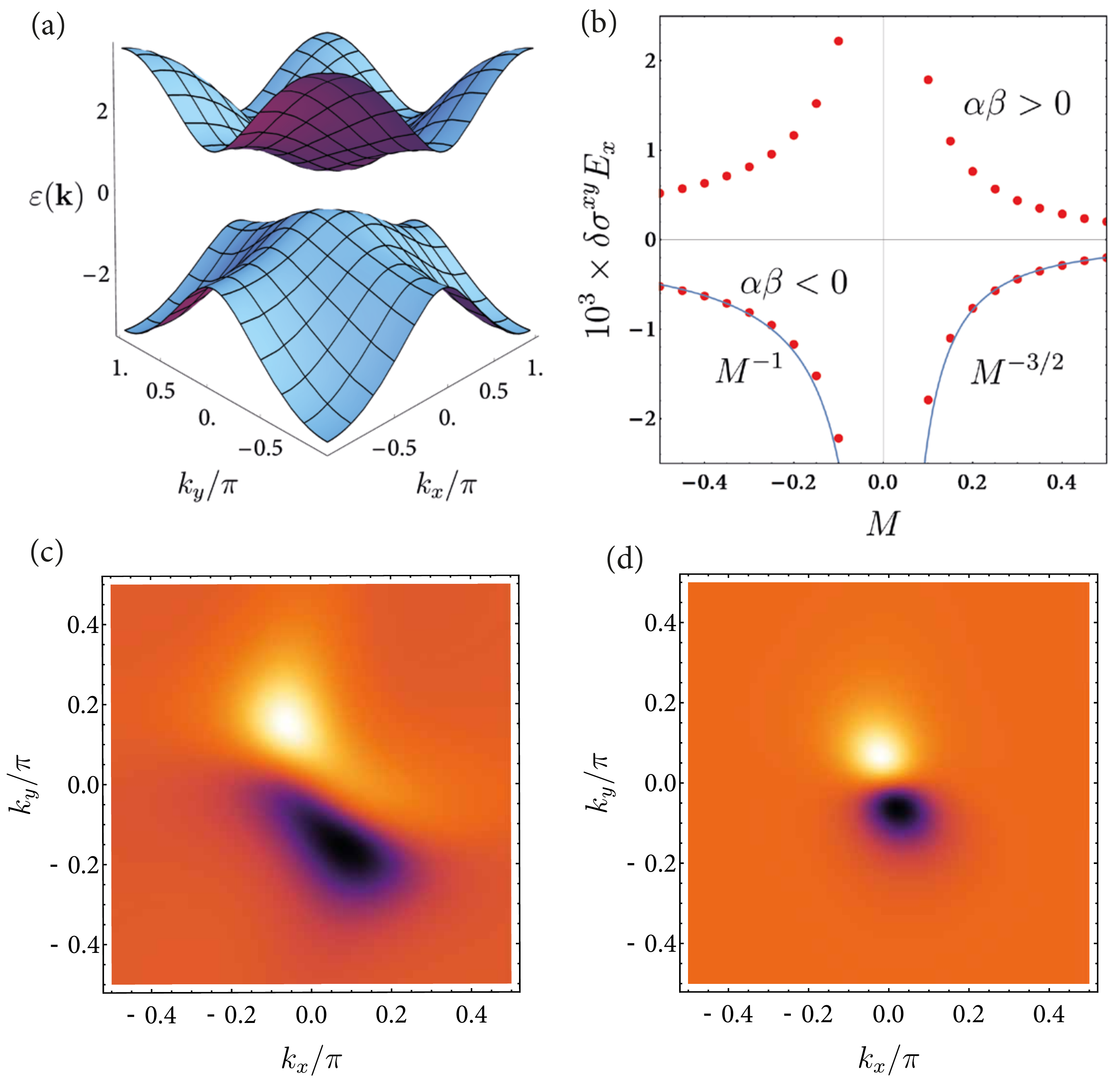}
    \caption{Nonlinear correction to the anomalous Hall conductivity of a gapped Dirac cone. (a) A two band model (Eq.~\eqref{eq:model_ham}) in the topological phase, when $M = -0.5$. (b) Magnitude of the correction $\delta \sigma^{xy}$ to the Hall conductivity, for $E_x = 1$. The sign of the correction depends on the sign of the product $\alpha \beta$. In the topological phase ($M<0$), the correction decays like $M{-1}$, while in the trivial phase ($M>0$) it decreases with $M^{-3/2}$. (c-d) Momentum space distribution of $I_1-I_3$ (Eqs.~\eqref{eq:i1}-\eqref{eq:fconductivity}) for the topological phase (c) and the trivial phase (d)}
    \label{fig:fig1}
\end{figure}
The nonlinearities $L_x, L_y$ do not merely break inversion and mirror symmetries remaining mirror but add higher derivatives of the Hamiltonian, i.e., $w^{ab} = \partial_a \partial_b H_0$, and induce a correction to the current operator, according to Eq.~\eqref{eq:single_particle_curr}. In the absence of non-linearities, sum rules \cite{Aversa1995}, notably that the fact that $\partial_a v^b_{nm} = i[r^a,v^b]_{nm} + iv^b_{nm} \delta^a_{nm} + ir^a_{nm}\Delta^b_{nm}$ emerge, enforcing cancellations between terms and nulling the correction. In Fig.~\ref{fig:fig1}(b), we show the magnitude of the correction as a function of of the mass parameter $M$ of the model. The response diverges as $M \to 0$ due to the presence of $\varepsilon_{nm}^{-n}$, $n>1$ in all terms. Unlike the Berry curvature, the flux of this expression, i.e. the integral $\sim \int \textrm{d}k \varepsilon^{-n} k$ does not equal a constant but decays as $k^{-3}$ in the leading order. This makes the correction fundamentally different from the Berry curvature, as it is highly singular in the limit $k \to 0$ for a vanishing mass. The corrections are sensitive to the topology of the system, as shown in Fig.~\ref{fig:fig1}(b). In the non-topological case ($M>0$), with vanishing total Berry phase, the corrections decay roughly as $M^{-3/2}$ while for $M<0$, the decrease is of the form $M^{-1}$, indicating a slower suppression. The momentum space distribution is also different for the two cases. In Figs.~\ref{fig:fig1}(c-d) we plot the momentum space density of the terms $I_1-I_3$ (Eqs.~\eqref{eq:i1}-\eqref{eq:fconductivity}). In the trivial case ($M>0$,   Fig.~\ref{fig:fig1}(d)), the divergence around the $\Gamma$ point of the model is clearly apparent, while when $M < 0$ the density is diffused across a broader region in momentum space, with the peak way from the $\Gamma$ point. As stated above, in the 2-band limit, $I_3$ (Eq.~\eqref{eq:fconductivity}) cannot contribute, and the non-linearity is most significantly encountered for the velocity shift $I_1$ (Eq.~\eqref{eq:i1}). 

\end{document}